# Heterogeneous Computing Systems

Dimple P. Khatri

dkhatri1@umbc.edu

University of Maryland, Baltimore County

Guanqun Song

Song.2107@osu.edu

The Ohio State University

Ting Zhu

zhu.3445@osu.edu

The Ohio State University

*Abstract*— **This survey of heterogeneous computing systems will help in analyzing the technological trends that will be at the basis of heterogeneous computing systems, highlighting the major opportunities and challenges such technologies will bring with them. This will help to understand the importance of the heterogeneous computing systems, which are becoming common architectural elements of not only the modern data centers but also highly integrated devices (IoT). Identify problems related to it, such as the resource allocation problem, middleware, processing architectures, programming challenges, etc. from the perspective of heterogeneous resources.**

## I. Introduction

Machine learning (ML) and deep learning (DL) algorithms are emerging as the new driving force behind the evolution of computer architecture. With the increasing adoption of ML/DL techniques in Cloud and high-performance computing (HPC) domains, several new architectures ranging from chips to entire systems have been introduced to the market to better support ML/DL-based applications. While HPC and Cloud have long been distinct domains with their own set of challenges, an increasing number of new applications are pushing for their rapid convergence. Many accelerators, such as GP-GPUs, FPGAs, and customized ASICs with dedicated functionalities, such as Google TPUs and Intel Neural Network Processor, are being developed, further expanding the data center heterogeneity landscape.

Today's Internet of Things (IoT) research has been very comprehensive, focusing on security [1-2], low-power communication [3], heterogeneous and concurrent communication [4-10], etc. In addition, IoT devices began incorporating specific acceleration functions, all to conserve energy. Outside of ML/DL applications, application acceleration is common; here, scientific applications popularized the use of GP-GPUs, as well as other architectures such as Accelerated Processing Units (APUs), Digital Signal Processors (DSPs), and many-cores. On the one hand, training complex deep learning models necessitates powerful architectures capable of performing a large number of operations per second while consuming minimal power; on the other hand, flexibility in supporting the execution of a broader range of applications is still required. From this perspective, heterogeneity is also pushed down: chip architectures include a mix of general-purpose cores and dedicated accelerating functions.

Supporting such large-scale heterogeneity necessitates an adequate software environment capable of maximizing productivity and extracting maximum performance from the underlying hardware. When looking at single platforms, Such as Nvidia's CUDA programming framework for supporting a flexible programming environment, OpenCL is a vendor-independent solution targeting different devices ranging from GPUs to FPGAs [11-12]. However, moving at scale and effectively exploiting heterogeneity remains a challenge. New challenges arise when such a large number of heterogeneous resources of varying types need to be managed. Most tools and frameworks for managing the allocation of resources to process jobs still offer only limited support for heterogeneous hardware.

Management and programmability will be difficult with such a wide diversity of hardware pieces. Memories, storage, and connectivity will all become more heterogeneous in the future. Operating systems, orchestration tools that are commonly used to manage large-scale infrastructures, and middleware will be obliged to adjust their resource assignment policies and schemes to accommodate for hardware variety in this situation. Likewise, programming languages will need to incorporate more sophisticated algorithms to efficiently utilize underlying hardware and spare programmers from manually targeting different architectures, allowing them to concentrate on the applications. As hardware evolves and becomes more distinct, apps must be rethought from the ground up to effectively utilize the novel features and reap the benefits. Development frameworks must adapt to the changes to assist programmers in making use of new hardware. To this objective, programming languages and development frameworks will expand to include more and more features, removing the need for programmers to elaborate on how to achieve a certain goal.

The possibilities of heterogeneous computing systems and IoT may bring us closer to digital solutions that would have been unimaginable just a few decades ago. By using heterogeneous computing systems and IoT, tasks that require enormous processing power, such as artificial intelligence research, can now be done cost-effectively and within tight timelines. Other potential applications of heterogeneous computing include medical imaging, remote sensing, pattern recognition, robotics, analytics, cybersecurity, and cross-protocol communications [13-22]. In summary, heterogeneous computing systems combined with wireless communications open up many new opportunities for businesses and researchers in the modern technological era.

Despite the continual advancements in digital computer manufacture, there is always an increasing demand for additional performance and capabilities. Such demand stems from a variety of sources, including end-user apps that require greater capabilities and efficiency to run longer on batteries, Cloud providers that must process an increasing huge quantity of data [23], and scientific applications that demand best-in-class performance. Heterogeneity has been widely used to satisfy this insatiable computing power demand while also being able to confront all of the obstacles that come with it. Heterogeneity has grown pervasive and multidimensional, spanning chip designs, nodes, and distributed systems [24-29]. Furthermore, achieving such objectives necessitates a more comprehensive strategy that incorporates several technologies.

## II. LARGE-SCALE, DISTRIBUTED AND HETEROGENEOUS ENVIRONMENTS REQUIREMENTS

Even if it's merely for increased capabilities, performance is critical for any modern application: resource requirements are continually increasing. Unlike in the past, when application developers simply had to wait for the following CPU generation, which was usually much faster, current processors barely improve in terms of clock speed. Instead, as the number of cores grows, new ways of organizing and designing applications are required. As a result of the expanded distribution, applications have long since departed the shared memory realm and must now cater to explicit communication and synchronization as well.

### A. Processing Architectures

Aside from GPPs' historical growth from single-core in-order microarchitecture to modern out-of-order multi-cores, developing PAs have seen widespread acceptance in recent years. GPUs have evolved into increasingly complicated systems after being designed to allow the viewing of complex 2D/3D scenes. GPUs continue to be based on micro-architectural designs with thousands of simple efficient cores, but they now include a growing number of specialized ones. With advancements in manufacturing technology, vast numbers of transistors could be integrated fast, allowing GPUs to become better suited to scientific applications. Indeed, GPUs became the favored platform for numerous generations of supercomputers due to the availability of enormous numbers of parallel cores optimized for conducting high-throughput mathematical computations.

The recent proliferation of ML/DL-based applications, as well as the growing demand for real-time realistic picture rendering, has brought new difficulties and adjustments to GPU architectures. Dedicated cores have been developed to better serve ML/DL workloads: the hardware structure has been modified to conduct matrix multiply and accumulate operations in a few clock cycles. For embedded graphic cores, the same design adjustments were made. Dedicated ray-tracing cores, such as those found in the Nvidia core-RT architecture, have only lately been released.

Continuous advancements in the shrinking of electronic chips are accompanied with a decrease in the reliability of the created gadgets. Transistor shrinkage increases the likelihood of malfunctioning components. This is especially true with today's production methods, where reaching acceptable yields has become more difficult. Controlling all the impacting parameters gets difficult when transistor sizes are reduced. The smaller the transistor, the more likely random fluctuations will alter its characteristics, producing misbehavior in some instances. Much However, transistors with such small feature sizes do not have the same (electrical) characteristics, making the creation of complicated systems even more challenging.

Future modular systems are predicted to make extensive use of technology integration, making erroneous intended system behaviors even more likely. To deal with such uncertainty, future systems will need to incorporate monitoring/correction techniques that range from advanced error correcting codes to more complicated and intelligent integrated testing features. It is projected that the complexity of programs would increase as more heterogeneous computing resources become available. As enormous multi-threading systems become more widely available, current programming models grow more prone to failure. Indeed, expressing and correctly mapping parallelism on such systems is more complex for the programmer. One of the next difficulties will be to figure out how to allow compilers to extract parallelism in various ways.

### B. Self Reconfiguration Heterogeneous

Hardware accelerators, such as GPGPUs and FPGAs, nevertheless require power-saving strategies such as Dynamic Voltage and Frequency Scaling, as well as partial reconfiguration for FPGAs, to keep power usage in check. Many approaches to energy/power optimization and adaptation focus on the hardware level, such as task scheduling combined with GPU-specific DVFS and dynamic resource sleep (DRS) algorithms to reduce total energy use.

## III. HETEROGENEOUS PROGRAMMING MODEL

Heterogeneous computing systems often have one or more CPUs, each with its own set of computer cores, as well as a graphics processing unit (GPU). Furthermore, data centers are increasingly integrating heterogeneous hardware into their servers, such as Field-Programmable Gate Arrays (FPGAs), allowing their clients to accelerate their applications using programmable hardware [29-32].

### A. Directive-Based Programming Models

Annotations in the source code, often known as pragmas, are used in directive-based programming. Directives are used to add annotations to existing sequential source code without having to change it. Users can run the program sequentially by simply ignoring the directives, or they can use a compatible compiler that can comprehend the directives and potentially build a parallel version of the original sequential code. These new directives tell the compiler where potentially parallel areas of code are located, how variables are accessed in different parts of the code, and how synchronization points should be handled.

*1) OpenACC:* The OpenACC (Open Accelerators) standard was proposed at the Supercomputing conference in 2011 after the success of OpenMP. This new standard uses a directive-based approach to allow programmers to create programs that run on heterogeneous architectures like multi-core CPUs and GPUs. Setting the range of the array that a kernel access is required by OpenACC. As long as the two sections must not overlap, this effectively allows separate kernels to access different parts of the same array without requiring synchronization.

*2) OpenMP:* Following the success of OpenACC for GPU-based heterogeneous programming, OpenMP added heterogeneous device support. This version of the standard adds a series of new directives for programming multi-core CPUs and GPUs. OpenMP is a lot more verbose than OpenACC when it comes to GPU programming. This is because OpenMP specifies how to use the GPU right down to the program directive level. In addition, OpenMP 4.0 offered new vocabulary for expanding the current OpenMP for multi-cores and shared computing memory systems to express computation for GPUs.

Both OpenMP and OpenACC are parallel programming libraries for C/C++ and Fortran programs, and they demand parallel programming skills as well as a thorough grasp of the target architecture to achieve performance. Programmers require a strong understanding of the data flow in their application to design heterogeneous applications with OpenMP and OpenACC. They must also comprehend the many barriers (both explicit and implicit), how to accomplish synchronization, share variables, how data is transported from CPU to GPU and vice versa, and how execution is mapped to the tar- get architecture. Furthermore, despite targeting heterogeneous architectures, and more specifically GPUs, these two models are not interchangeable. OpenACC was created from the ground up to extend the OpenMP standard, which targets multi-core CPUs.

Some programmers rebuild OpenMP programs that target multi-core CPUs using OpenACC directives in order to adapt their applications to heterogeneous devices with the least amount of effort possible. However, given the overheads of data transfers to and from the target device, as well as the limits imposed by GPU cores, simply converting OpenMP annotated applications targeting multi-core CPUs to OpenACC may not provide in performance gains. On the contrary, it's possible that it'll end up performing worse than sequential execution.

### B. Explicit Heterogeneous Programming Models

In comparison to directive-based heterogeneous programming, such as OpenMP, explicit heterogeneous programming is relatively recent [33-35].

*1) OpenCL:* The Khronos Group published OpenCL (Open Computing Language) as a heterogeneous computing standard in 2008. The OpenCL standard is divided into four sections: Platform model, Execution model, Programming model and Memory model. The host and device model in which OpenCL programs are organized is defined by the platform model. The execution model specifies how programs are executed on the target device and how the host manages this process. It also defines OpenCL kernels, which are functions that must be run on the devices, as well as how they must be synchronized. The programming model specifies the data and task programming paradigms, as well as how the accelerator and host are synchronized. The OpenCL memory hierarchy and OpenCL memory objects are defined by the memory model.

*2) CUDA:* The Common Unified Device Architecture (CUDA) is a parallel programming framework and language that allows C/C++ programs to run on NVIDIA GPUs. CUDA and OpenCL share a lot of the same principles, vocabulary, and ideas. In fact, the CUDA programming language was the primary inspiration for OpenCL. The distinction is that because it can only run on NVIDIA technology, CUDA is slightly simpler and better streamlined. A CUDA application is made up of two parts: host code, which is code that runs on the main CPU, and device code, which is the function that runs on the GPU and is accelerated. CUDA programs employ streams to transfer commands between the host and devices, similar to OpenCL. This design is almost identical to the view of an OpenCL program. Physical memory is built into the main CPU. Apart from that, the GPU has its own physical memory. The following is a typical CUDA application workflow: Set up the host buffers and the device buffers. Then to copy the data on the device, the CPU sends a data transfer from the host to the device (HD). Now a CUDA kernel is launched by the CPU. It specifies the number of CUDA threads to be used. Finally the CPU sends data from the device to the host (DH) so that the results can be copied to the main CPU.

CUDA and OpenCL have a lot in common. Both programs use the same programming model and essential concepts. Furthermore, each of these tasks necessitate a high level of programming ability. To understand and build applications with such models, a thorough understanding of the GPU architecture as well as the programming model is required. CUDA, on the other hand, is much easier to get started with because it tailors its model to NVIDIA GPUs. Furthermore, because it is specifically built for NVIDIA GPUs, CUDA is supposed to provide substantially higher performance. CUDA also provides many layers of abstraction, including high-level libraries, CUDA syntax, and the CUDA driver. This makes CUDA programming more adaptable and acceptable for a wide range of sectors and academia. OpenCL, on the other hand, is more portable, as it can run code on any OpenCL-compatible device, such as CPUs, GPUs from many vendors, and FPGAs (e.g., from Xilinx and Intel Altera).

## IV. REAL-TIME TASK MODELS FOR HETEROGENEOUS COMPUTING

The principal objective of real-time systems is not just that the output is correct, but that it is provided at the correct moment, which means that once a stimulus is received, the system's response time must be restricted by construction. This is usually accomplished by having a complete understanding of both software and hardware system components, which are thoroughly analyzed to determine their worst-case response time, or the time it takes to provide their output.

### A. Job-Based Models

*1) Periodic, Aperiodic, Sporadic:* Periodic tasks are made up of an infinite number of similar actions, known as instances or jobs, that are activated at a set rate on a regular basis. Aperiodic tasks are made up of an endless number of identical jobs that are not interleaved in a predictable way. Sporadic tasks are aperiodic tasks in which jobs are separated by a minimal inter-arrival interval.

*2) Multiframe Model:* In this approach, frame deadlines are allowed to diverge from the minimum separation; also, all frames do not have to have the same deadlines, and all minimum separations do not have to be the same.

*3) Elastic Model:* The concept behind this model is to treat each task as flexible, with a stiffness coefficient and length limits. The computation, in particular, is fixed, whereas the duration can be modified within a specific range.

*4) Mixed-Criticality Model:* We might assume that there is additional parameter to consider in the preceding models: the job's priority, which specifies its importance. However, in some cases, we are more concerned with the task's relevance and criticality (i.e., whether it is a hard or soft task). In this example, a model is concerned with taking into consideration various levels of criticality.

*5) Splitted Task:* For systems with minimal preemption, the splitted-task approach is recommended. Each task is divided into non-preemptive pieces (sub-jobs) in this approach, which is achieved by introducing preemption points into the code. Preemptions are only possible at the intersections of sub-jobs. All of the jobs created by a single task are divided into the same sub-job division.

*B. Graph-Based Models*

These models are uncomplicated and do not account for aspects such as parallelism modeling within individual jobs or precedence limits. Because there was no possibility of parallelism on a single processor, this was not a concern in the context of uniprocessor real-time systems. However, real- time systems are increasingly being implemented on multiprocessor, multicore platforms, necessitating the development of models capable of revealing any feasible parallelism within the workload, including precedence dependencies across various sections of the same task.

*1) DAG Model:* The sporadic directed acyclic graph (DAG) model is a broad parallel model that deals with the fine-grained execution provided by current parallel programming paradigms. Tasks are represented in this model by directed acyclic graphs with a single source vertex and a single sink vertex. The edges of the DAG indicate precedence restrictions between these jobs, while the vertex of the DAG represents a sequential job.

*2) Digraph Model*: There is a more general DAG model called digraph real-time (DRT) job model that employs only directed graph. A directed graph G(T) describes a DRT job T, with edge and vertex labels as before. Any directed graph can be used to describe a task; there are no more limitations. Every cycle in that model, on the other hand, must pass via the source vertex.

V. SECURITY FOR HETEROGENEOUS COMPUTING

Heterogeneous systems, in contrast to homogeneous systems, necessitate customization and specific configuration by their very nature. These variances may necessitate independent verification and certification for each variant, which would not only increase deployment costs but also introduce subtle deviations that an intruder may exploit. As a result, we require common procedures and mechanisms to ensure that heterogeneous systems have the same level of security as homogeneous systems. Furthermore, heterogeneous systems are more likely to withstand common- mode attacks, which use a single effective attack tactic to compromise the entire population.

Furthermore, heterogeneity can extend beyond a single system, forming a system-of-systems with several systems, each with its management environment. To take advantage of the total system's capabilities, several configurations and software drivers will need to be integrated. Although there may be some tool overlap, older computers may run older applications for compatibility reasons. This necessitates the hiring of general administrators, who may face a high learning curve as they try to become acquainted with the super-system.

When creating applications for such heterogeneous systems, new considerations must be made in comparison to homogeneous systems. Differentiation is introduced by multiple sorts of computer units in the same system, which may include different instruction set architectures, different memory layouts, variances in available libraries/OS services, different interconnects, and varying performance, among other things.

A heterogeneous system is only as secure as its weakest component from a security standpoint. If a vulnerable component is compromised, an attack could infect the entire system. Each component of the overall system may come from a different source, with its own set of security features, communication protocols, and system updates. If a part is offered as a black box, such as with proprietary licensing, it may do actions that aren't intended, which could be malicious.

However, heterogeneous systems face numerous problems, making the development of such methods difficult. Heterogeneous systems use a variety of communication technologies, making it difficult to build secure connections using multiple technologies. Furthermore, the computational power of the nodes in the heterogeneous system may differ. As a result, not all proposed secure protocols could be implemented. Furthermore, in a heterogeneous system, several security policies coexist and must be negotiated in order to arrive at a single generic security policy for the entire system.

Creating this generic policy is difficult because it necessitates a thorough understanding of all conceivable communication pathways between all system components. Because of the system borders among the various vendors, such knowledge is difficult to obtain. Another area that can be examined in order to detect potential assaults is the behavior of the network/communication bus. A Network Intrusion Detection System (NIDS) is a system that examines incoming network traffic for unusual payloads before they reach a critical system. The inspection is carried out by comparing incoming traffic to patterns known to be present in malicious packets.

*Four categories that pose security issues in a heterogeneous systems are:*

*1) Security requirements:* Components from various vendors have varying levels of security in a heterogeneous system. Sets of components are developed by various teams with varied levels of expertise. Even if each of these sets has been constructed in a secure manner, combining them in the same system may result in the system becoming insecure. In a vehicle system, for example, an attacker can benefit from the heterogeneity in the security levels of the various ECUs by launching a stepping stone attack across the entire system. Attackers begin by compromising weak components or subsystems and using them as an attack surface to wreak havoc on all associated subsystems.

*2)* Communication: Different vendors' parts support a variety of communication protocols. However, in a certain design, these components must interact transparently with one another in order for the entire system to function successfully. This necessitates continuous communication between these components during their integration and operation. However, because it must adhere to the needs of all components, such a procedure introduces additional security flaws.

*3) Management:* The management of a heterogeneous system's resources is another crucial feature. Administrators must coordinate all previous parts of the system (security, communication, and updates) so that they have a complete view of how it works and can intervene if something goes wrong. Furthermore, heterogeneity can extend beyond a single system, becoming a super-system of numerous systems, each with its own management environment. To take advantage of the total system's capabilities, several configurations and software drivers will need to be integrated. Although there may be some tool overlap, older computers may run older applications for compatibility reasons. This necessitates the hiring of general administrators, who may have a steep learning curve in becoming acquainted with the super-system.

*4) Updates:* Persistent protection means that the system software and defensive mechanisms are kept up to date with new threats. For example, all equipment in a smart home/building must be secured and updated with the newest firmware/patches at all times. This requirement for regular updates raises an intriguing dilemma, notably the necessity for security updates against availability. On the other hand, updating a gadget always runs the danger of leaving it "bricked," meaning it's no longer functioning and unlikely to work again. Furthermore, updating one component may need updating other components as a result. If one component fails to be updated, the system may become unusable. In this situation, rolling back the upgraded component to restore functionality could lead to a number of security flaws.

Above requirements shows that there is need to develop techniques that promote security for heterogeneous systems starting from prevention, detection and response. It is possible to monitor the activity of a component or job without incurring significant overhead, but this requires effort throughout the development process. Major components of what will eventually become the system security policy must be defined as the design progresses by employees who are intimately familiar with each component's design. Security is thus incorporated into the architecture rather than being an afterthought. The trend toward complex systems with internal networks of heterogeneous computing resources is expected to continue because it allows designers to select the most appropriate and cost-effective elements for their designs. Furthermore, because these systems are embedded in expensive platforms (like as automobiles), they must be capable for an extended period of time.

## Conclusion

Digital technologies are changing our lives and how we interact with our surroundings. Modern IoT devices make it possible to create a digital representation of real-world phenomena, making it easier to understand them. The other side of the medal represents the massive amount of data that needs to be analyzed to extract information and obtain a clear understanding of the phenomenon. Machine learning and deep learning techniques are changing our way of analyzing this massive amount of data. In this context, heterogeneity has become critical for effectively supporting their execution across infrastructures and platforms that span the boundaries of single systems and data centers. Future advancements in technology will amplify such diversity in hardware and software components, presenting new challenges.